\documentclass[a4paper]{scrartcl}
\usepackage{a4wide}
\usepackage[T1]{fontenc}
\usepackage[utf8]{inputenc}
\usepackage{amsmath, mathtools, amsfonts, amssymb, amsthm}
\usepackage{bm}
\usepackage{tikz}
\usepackage{float}
\usepackage{lmodern}
\usepackage{fix-cm}
\usepackage{hyperref}
\usepackage{subfig}

\newcommand{\OPT}{\ensuremath{\operatorname{\Omega}}}
\newcommand{\UB}{\ensuremath{M}}
\newcommand{\LB}{\ensuremath{\mu}}
\newcommand{\R}{\mathbb{R}}  
\renewcommand{\bar}[1]{\overline{#1}}

\tikzstyle{close} = [fill = black, draw = black, circle, scale = 0.4]
\tikzstyle{open} = [draw = black, circle, scale = 0.4]
\tikzstyle{pure} = [fill = red, draw = black, rectangle, inner sep=0.7mm]
\tikzstyle{help} = [above, sloped]

\newtheorem{theorem}{Theorem}
\newtheorem{corollary}[theorem]{Corollary}
\newtheorem{lemma}[theorem]{Lemma}

\newtheorem{definition}[theorem]{Definition}

\newtheorem{remark}[theorem]{Remark}

\begin{document}

\sloppy

\title{Energy-efficient Routing of Hybrid Vehicles}
\author{Christian Schwan \phantom{and} Martin Strehler\\ \small Brandenburg University of Technology\\ \small Post office box 10 13 44\\ \small 03013 Cottbus, Germany\\ \small \url{christian.schwan@b-tu.de}\quad\url{martin.strehler@b-tu.de}}

\maketitle

\abstract{We consider a constrained shortest path problem with two resources. These two resources can be converted into each other in a particular manner.
Our practical application is the  energy optimal routing of hybrid vehicles. Due to the possibility of converting fuel into electric energy this setting adds new characteristics and new combinatorial possibilities to the common constrained shortest path problem (CSP). We formulate the resulting problem as a generalization of CSP. We show that optimal paths in this model may contain cycles and we state conditions to prevent them. The main contribution is a polynomial-time approximation scheme and a simpler approximation algorithm for computing energy-optimal paths in graphs.}

\section{Introduction}

Hybrid vehicles are becoming a more and more attractive alternative for resource-efficient individual travelling. Moreover, interesting routing problems arise in this context. When should the classical fuel-powered engine be used and when the electric one? When should the battery be charged? Which route towards the destination should be used to minimize overall fuel consumption? Does the route choice depend on the the initial state of the battery?

As a first step to answer these questions, we study a simplified model of an \textit{autonomous parallel} hybrid vehicle and formulate a corresponding constrained shortest path problem. 

There are various different types of hybrid cars. The most commonly produced hybrid vehicles are at present parallel hybrid systems.
 Such a vehicle possesses a combustion engine and an electric motor that are joined at a common axis in parallel and they are powering the car together. A battery is supplying the electric motor with electricity. 
In return, using the electric motor as a generator, the battery can be charged in two ways. Firstly, energy may be recuperated by using the brake of the car or by driving downhill. Secondly, the combustion engine may provide additional torque to power the generator. In other words, we can use additional fuel to generate electric energy. 

Linking engine control and routing offers a great potential for saving energy. Exact knowledge of a route may be used to find an optimal control of both the electric and the combustion engine. Using the electric motor/generator can shift the load curve in the fuel-consumption map to a better operating point of the combustion engine. Knowing topography and the desired speed in advance an optimal engine control strategy can be calculated.  

On the other side, if a given engine control takes route information into account, fuel consumption on a certain road depends on the whole route. That is, two routes with different origins and destinations may use the same road, but the individual control strategies may differ significantly on this common road. Therefore, also routing has to anticipate the behaviour of the engine control to find energy optimal routes.

In this paper, we aim to find an energy-optimal routing of hybrid vehicles. More precisely, we want to compute a path from origin to destination with minimum fuel consumption, i.e., minimizing fuel is our objective. However, given an initial state of the battery, we must not exceed certain lower and upper bounds on the battery charge. Thus, the battery introduces additional resource constraints. As we will show, a shortest---that is, an energy-optimal---path can contain cycles. While this may be surprising at the beginning, our examples will show that such cycles can be caused already by small differences in the efficiency of the different resources and their conversion. Since cycles are most likely not welcome in practice, we discuss sufficient conditions to prevent such cycles. We present both a more theoretical polynomial time approximation scheme and an application oriented approximation algorithm to tackle the problem. 

\section{Related Work}

Constrained shortest paths with independent resources have been studied for a long time. The original problem is well known to be $\cal NP$-complete which can be easily shown by reduction of \textsc{Partition} or \textsc{Knapsack}~\cite{Johnson_complexity}. Several approximation methods have been developed.  An algorithm by Warburton~\cite{war87} for listing Pareto-optimal multi-objective shortest paths is used by Hassin~\cite{Hassin} to develop an approximation algorithm based on rounding and scaling. Ziegelmann and Mehlhorn~\cite{Ziegelmann+Mehlhorn_CSP, Ziegelmann_CSP} deal with the dual formulation of the path-based CSP. There, the authors are iteratively computing shortest paths and updating the new objective function which is a linear combination of the different costs and resources. Recently, Garcia~\cite{Garcia_CSP} studied the integer programming formulation and cutting planes for CSP. A survey on different approaches can also be found in \cite{Garcia_CSP} and \cite{Ziegelmann_CSP}.

Hybrid cars are rarely studied in this context. In~\cite{elektro:modellierung}, the authors focus on pure electric vehicles and recuperation. In their approach, after correcting energy consumption by potential energy, strictly positive costs are achieved and an $A^*$ algorithm can be applied. However, no other parameters are considered. Thus, the problem reduces to a simple shortest path problem. 

Additionally, optimal control of the powertrain is the essential counterpart in our application. Already the control of a hybrid vehicle itself offers potential for saving fuel~\cite{hybrid:Zhang:motorsteuerung}. Using road information in the control of a hybrid car is suggested by Back in~\cite{back:motorsteuerung}. Larsson et al.~\cite{hybrid:Larsson:route+steuerung,hybrid:Larsson+:route+steuerung} develop a hybrid powertrain control which uses precomputed optimal strategies. Therefore, they identify frequently travelled routes from logged driving data, i.e., optimal control for fixed routes is computed. In practical experiments, up to 10\% of fuel could be saved on these routes compared to standard depleting-sustaining strategies.

\section{Preliminaries}\label{sec:preliminaries}

A road network is modelled as a finite directed graph $G=(V,E)$ with $|V|=n$ vertices and $|E|=m$ edges. Additionally, we consider two resources, namely fuel and battery. However, resource consumption on a given edge is not fixed. We may use the upcoming road for charging the battery, fully support the combustion engine with the electric motor, or anything in between.

Before stating the complete description of our problem in the end of this section, let us study resource consumption more closely. The results in this paper come from a research project funded by the German Federal Ministry of Education and Research (grant number 05M13ICA). The parties hereto are mathematicians and industrial partners from the automotive and navigation sector. The methods used include both discrete optimization and optimal control.

Realistic resource consumptions are derived by a model predictive control approach based on a sophisticated engine model, road data and topographical information. Here, we compute all Pareto-optimal control strategies with respect to fuel and electrical charge consumption for each road. The optimal strategy to be used on a specific road has to be chosen by the routing algorithm. Therefore, we introduce the strategy parameter $\alpha_e\in [0,1]$ for $e\in E$. Here, $\alpha_e=1$ means choosing the control strategy with maximal minimum fuel consumption and charging the battery. Contrary, $\alpha_e=0$ corresponds to the control strategy with minimal minimum fuel consumption and depleting the battery. Furthermore, we use a realistic battery model, i.e., fuel consumption on an edge also depends on the initial battery charge.

Since a complete description of this approach would be beyond the scope of this paper, we use a simplified model here. The resource consumption is parameterized by $\alpha$ such that we obtain two functions
\begin{align}
c_{f,e}:&[0,1]  \to \R_{\ge 0}\mbox{ and}\\
c_{b,e}:&[0,1]  \to \R
\end{align}
for every edge $e\in E$ where $c_{f,e}$ and $c_{b,e}$ map the strategy parameter $\alpha$ to the actual consumption of fuel and battery, respectively. In accordance to physics, we require for every cycle and every choice of $\alpha$, that the total fuel consumption or the total battery consumption on this cycle is positive. We refer to functions fulfilling this property as \emph{conservative} resource consumption function.

For simplicity, we may choose one of the functions to be affine in $\alpha$. Besides the path itself, the routing algorithm also has to choose the parameter $\alpha$.

\begin{definition}
 An $s$-$t$-\emph{path} $P$ is a sequence of ordered pairs $(e_i,\alpha_{i}), i\in\{1,\dots,k_P\}$ such that the $e_i\in E$ fit head to tail, $e_1$ starts in $s\in V$ and $e_{k_P}$ ends in $t\in V$. 
\end{definition}

An edge may appear several times in a path\footnote{More formally, one should name such an object a \emph{walk}. However, we use the term \emph{path} here to indicate that we are going to compute paths eventually.}, i.e., $e_i=e_j$ for $i\not=j$ is possible. In this case, the corresponding $\alpha_{i}$ and $\alpha_{j}$ may also differ. For each $s$-$t$-path $P$, the intermediate battery states are computed by $B_i=B_{i-1}-c_{b,e_i}(\alpha_{i})$, $i\in\{1,\dots,k_P\}$ where $B_0$ is the initial state of the battery in $s$. All battery charges have to obey the capacity constraints, i.e., the charge always has to be within zero and the battery capacity $\overline{B}$.

\begin{definition}
An $s$-$t$-path $P$ is called \emph{feasible}, if it satisfies the constraints
\begin{align}
0\le B_i =  B_{i-1}-c_{b,e_i}(\alpha_{i}) \le\bar{B} \label{eq:constrained-1} \qquad \text{for all } i=1,\ldots,k_P\;.
\end{align}
\end{definition}

Note that the parameter $\alpha_i$ can be interpreted as a general strategy parameter that suggests the final state of the battery at the end of the upcoming road segment. For example, optimal control ensures that battery bounds are not exceeded within an edge.

Now, the cost of $P$ is $\sum_{i=1}^{k_P} c_{f,e_i}(\alpha_{i})$. Let $\mathcal{P}$ be the set of all paths $P$ from start node $s$ to target node $t$ and let $k_P$ denote the number of edges in $P$. 
Using binary decision variables $x_P$ for each path $P$ to select this particular path, a condensed version of our problem reads as follows:

\begin{subequations}\footnotesize
\begin{align}
 \bm{(P)}\qquad \min_{x_P,\alpha} \qquad &\sum_{P\in\mathcal{P}} \left\lbrace  \sum_{i=1}^{k_P} c_{f,e_i}(\alpha_i) \right\rbrace x_P \label{eq:problem:objective}  \\
 \operatorname{s.t.} \qquad & 0\le B_0 - x_P\sum_{i=1}^{j} c_{b,e_i}(\alpha_i)  \le \bar{B} \qquad  \forall P\in\mathcal{P},~ j=1,\dots,k_P \label{eq:problem:feasible}\\
 & \sum_{\mathcal{P}} x_P =1\\
 & x_P \in \{0,1\} \qquad \forall P \in \mathcal{P}\label{eq:problem:01-rest} \\
 & \alpha_i \in [0,1] \label{eq:problem:alpha-rest}
\end{align}
\end{subequations}

Before solving problem $\bm{(P)}$, we will study some important properties of the solutions and the problem's complexity in the next two sections.

\section{Cycles in Shortest Paths}\label{sec:complexity}

Although cycles in an energy-efficient path in a graph with conservative resource consumption functions sound very implausible at first, we will show in the following that they can emerge quite naturally from the consumption functions. For example, such cycles exist if the resource consumption blocks an edge while the battery is not yet completely charged. But also slight differences in the consumption functions may lead to cycles.

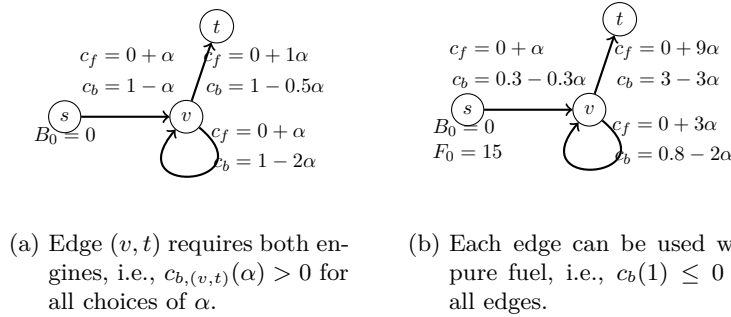
\begin{figure}[ht]\centering
\subfloat[Edge $(v,t)$ requires both engines, i.e., $c_{b,(v,t)}(\alpha)>0$ for all choices of $\alpha$.\label{fig:cycle1}]{
\begin{tikzpicture}[->, every node/.style={circle,draw, scale=0.7}, t/.style={align=right, draw=none}, scale=0.4]
  \node (1) at (0,0) {$s$};
  \node (2) at (4,0)  {$v$};
  \node (3) at (5,3) {$t$};
  \node[t, align=left] at (0,-1) {$B_0=0$\\
  };
 \draw[thick] (1) to 
 (2);
 \node[t] at (2,1.5) {$\begin{aligned}
                                       c_f&=0+\alpha\\
                                       c_b&=1-\alpha\\
                                      \end{aligned}$};
 \draw[thick] (2) to 
 (3);
 \node[t] at (6.5,1.5) {$\begin{aligned}
                                       c_f&=0+1\alpha\\
                                       c_b&=1-0.5\alpha\\
                         \end{aligned}
 $};
 \path[->, thick]      (2)  edge  [loop below, distance=3cm, out=320, in=230]  
 (2) ;
 \node[t] at (6.6,-1) {$\begin{aligned}
                                       c_f&=0+\alpha\\
                                       c_b&=1-2\alpha\\
                     \end{aligned}$
 };
 \end{tikzpicture}
}\qquad
\subfloat[Each edge can be used with pure fuel, i.e., $c_{b}(1)\le0$ for all edges.\label{fig:cycle2}]{
\begin{tikzpicture}[->, every node/.style={circle,draw, scale=0.7}, t/.style={align=right, draw=none}, scale=0.4]
  \node (1) at (0,0) {$s$};
  \node (2) at (4,0)  {$v$};
  \node (3) at (5,3) {$t$};
  \node[t, align=left] at (0,-1) {$B_0=0$\\$F_0=15$
  };
 \draw[thick] (1) to 
 (2);
 \node[t] at (1.6,1.5) {$\begin{aligned}
                                       c_f&=0+\alpha\\
                                       c_b&=0.3-0.3\alpha
                                      \end{aligned}$};
 \draw[thick] (2) to 
 (3);
 \node[t] at (6.5,1.5) {$\begin{aligned}
                                       c_f&=0+9\alpha\\
                                       c_b&=3-3\alpha
                         \end{aligned}
 $};
 \path[->, thick]      (2)  edge  [loop below, distance=3cm, out=320, in=230]  
 (2) ;
 \node[t] at (6.8,-1) {$\begin{aligned}
                                       c_f&=0+3\alpha\\
                                       c_b&=0.8-2\alpha
                     \end{aligned}$
 };
\end{tikzpicture}
}
\caption{Two instances of a network with different consumption functions.}\label{fig:cycle}
\end{figure}

Consider a certain road with an extremely steep increase that requires both the combustion engine and the electric motor to climb the hill, but the battery is empty. In other words, there is an edge $e$ with $c_{b,e}(\alpha)>0$ for all choices of $\alpha$. If the battery is empty, then passing this edge obviously yields an infeasible battery charge. The example in Figure~\ref{fig:cycle1} exploits this fact to enforce a cycle in every feasible path.

Consider the graph in Figure \ref{fig:cycle1} and let us start  in $s$ with an empty battery, e.g., $B_0=0$. We can only use $\alpha_1=1$ on $e_1=(s,v)$. All other choices would yield a negative value $B_1$. Now, we cannot pass $(v,t)$, since we need at least 0.5 units in the battery to do so ($\alpha=1$). Hence, we use the cycle $(v,v)$ with $\alpha_{2}=1$. This yields $B_2=0.5$, only afterwards $(v,t)$ can be traversed. Hence, each feasible path visits $v$ at least twice.

To avoid impassable edges due to temporary insufficient resources, we can, of course, assume an appropriate sizing of the combustion engine. That is, we require that each edge can be used with every initial battery charge, potentially causing high costs.

\begin{remark}[Cycle prevention condition~I] \label{remark:alpha0}
An instance fulfills the \textsc{Cycle prevention condition~I} if for each $e \in E$ there exists $\alpha^0_e\in[0,1]$ with $c_{b,e}(\alpha^0_e)= 0$.
\end{remark}

Consequently, if \textsc{Cycle prevention condition~I} is fulfilled, then for any given path from $s$ to $t$, the choice of $\alpha_i=\alpha^0_{e_i}$ makes it a feasible one. That is, if $s$ is connected to $t$, then there exist feasible paths without cycles.

\begin{figure}[ht]\centering
\subfloat[Pareto-optimal labels at node $v$ with no cycle (square dot) and up to four cycles using $(v,v)$.]{
\begin{tikzpicture}[scale=0.28]
\draw[help lines, color=gray!30] (0,0) grid (15,5);
\draw[->] (-0.2,0) -- (15.2,0) node[right] {$\scriptstyle F$};
\draw[->] (0,-0.2) -- (0,5.2) node[above] {$\scriptstyle B$};
\foreach \x in {0,2, ..., 14}
\draw (\x,-.1) -- (\x,.1) node[below=4pt] {$\scriptstyle\x$};
\foreach \y in {0,2, ...,4}
\draw (-.1,\y) -- (.1,\y) node[left=4pt] {$\scriptstyle\y$};

\node [close] (a) at (2,4.8) {};
\node [open] (b) at (3.8,3.6) {};
\draw (a) to node[help] {$\scriptstyle 4\times (v,v)$} (b);
\node [close] (0) at (5,3.6) {};
\node [open] (1) at (6.8,2.4) {};
\draw (0) to node[help] {$\scriptstyle 3\times (v,v)$} (1);
\node [close] (2) at (8,2.4) {};
\node [open] (3) at (9.8,1.2) {};
\draw (2) to node[help] {$\scriptstyle 2\times (v,v)$} (3);
\node [close] (4) at (11,1.2) {};
\node [open] (5) at (12.8,0) {};
\draw (4) to node[help] {$\scriptstyle 1\times (v,v)$} (5);
\node [pure] (6) at (14,0) {};
\end{tikzpicture}
}\qquad
\subfloat[Pareto-optimal labels of node $t$. Pure combustion drive (square dot) is dominated by using the cycle.]{
\begin{tikzpicture}[scale =0.28]
\draw[help lines, color=gray!30] (0,0) grid (15,5);
\draw[->] (-0.2,0) -- (15.2,0) node[right] {$\scriptstyle T$};
\draw[->] (0,-0.2) -- (0,5.2) node[above] {$\scriptstyle B$};
\foreach \x in {0,2, ..., 15}
\draw (\x,-.1) -- (\x,.1) node[below=4pt] {$\scriptstyle\x$};
\foreach \y in {0,2, ...,5}
\draw (-.1,\y) -- (.1,\y) node[left=4pt] {$\scriptstyle\y$};

\node [close] (0) at (0,2.26667) {};
\node [close] (1) at (5,0.6) {};
\draw (0) to node[help] {$\scriptstyle$} (1);
\node [close] (2) at (5,0.6) {};
\node [open] (3) at (5.6,0.2) {};
\draw (2) to node[help] {$\scriptstyle$} (3);
\node [close] (4) at (5.6,0.2) {};
\node [close] (5) at (6.2,0) {};
\draw (4) to node[help] {$\scriptstyle$} (5);

\node [pure] at (5,0) {};
\end{tikzpicture}
}
\caption{Pareto-optimal labels for the scenario in Figure~\ref{fig:cycle2}.}\label{fig:labels}
\end{figure}
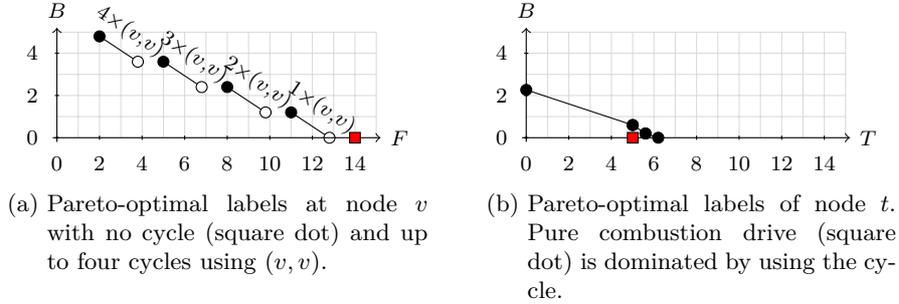

The instance in Figure~\ref{fig:cycle2} fulfills the \textsc{Cycle prevention condition~I}. Yet, the optimal path still contains cycles. To illustrate that let us fix some initial values, e.g., initial battery charge of $B_0=0$, battery capacity $\bar{B}=5$ and initial fuel budget $F_0=15$. For simplicity we normally assume infinite fuel resources, but choosing some bounded value here simplifies the presentation in Figure~\ref{fig:labels}. 

Consider Figure~\ref{fig:cycle2}; again, due to the empty battery the first edge forces $\alpha_{1}=1$. Now, we could pass $(v,t)$ directly using $\alpha_2=1$, which would cost additional 9 units of fuel. In total, $t$ is reached with fuel consumption of 10 units  and an empty battery. However, we may also use $(v,v)$ as the second edge $e_2$. Choosing $\alpha_2=1$, we charge the battery by 1.2 units paying 3 units of fuel. Thus, vertex $v$ is reached with 4 units of fuel consumption, which seems to be worse than before. But now, we can pass $(v,t)$ as third edge $e_3$ with $\alpha_3=0.6$. Hence, we need only 5.4 additional units of fuel, arriving at $t$ with a total fuel consumption of 9.4 units. Hence, one cycle saved 0.6 units of fuel. Even better, using the cycle twice with $\alpha_3=1$ at the second turn and $(v,t)$ as fourth edge with $\alpha_4=0.2$, we reach $t$ with only 8.8 units fuel consumption. All Pareto-optimal labels for all choices of $\alpha\in[0,1]$ are shown in Figure~\ref{fig:labels}.

Note that the total consumption on each edge of the graph in Figure~\ref{fig:cycle2} is positive.  That is, for every edge and every choice of $\alpha$ it holds $c_{f,e}(\alpha)+c_{b,e}(\alpha)>0$. The consumption functions do not contradict energy conservation, not even recuperation is used. In this example, cycles of the optimal solution  already emerge from different efficiency factors of the two engines.

Contrary to common shortest paths, where cycles with negative costs imply that a shortest $s$-$t$-path does not exist, positive total costs force to energy-optimal path to eventually reach $t$.

\begin{lemma}
If there exists a feasible $s$-$t$-paths and each cycle has positive total costs regardless of the choice of the $\alpha_{e_i}$, an energy-optimal $s$-$t$-path exists. That is, an optimal path may use an arbitrary but finite number of cycles.
\end{lemma}

Although cycles may exist in an optimal path in general, it is interesting what conditions guarantee that those cycles do not occur. This is relevant for understanding the underlying structure of this optimization problem. But maybe even more importantly, this is quite substantial for practical applications, where cycles seem to be rather hard to motivate. A sufficient, but not necessary condition is given in Theorem~\ref{theo:nocycles}.

\begin{theorem}[Cycle prevention condition~II]\label{theo:nocycles} Assume, \textsc{Cycle prevention condition~I} is fulfilled. If for every cycle $C=(e_1,\dots,e_k)$, every choice of $\alpha_{i}\in [0,1]$, $i\in\{1,\dots,k\}$, and each edge $\ell\in G$ with $\alpha_\ell\in [0,1]$ holds
\begin{align}
\frac{\sum_{i=1}^k c_{f,e_i} (\alpha_{i})}{\left\lvert \sum_{i=1}^k c_{b,e_i} (\alpha_{i}) \right\rvert} > -\frac{\frac{d}{d\alpha}c_{f,\ell}(\alpha_\ell)}{\frac{d}{d\alpha}c_{b,\ell}(\alpha_\ell)}\label{eq:cycleprevention2}, 
\end{align}
then the energy optimal path with lowest fuel consumption contains no cycle.
\end{theorem}

Both sides of the \textsc{Cycle prevention condition~II} describe the efficiency of exchanging resources. The left hand side describes the ratio of fuel consumption and battery gain on a cycle. The right hand side describes the local efficiency ratio on an edge. If the condition is not fulfilled, then it is potentially useful to use the cycle before passing this edge. Due to space constraints we omit the proof.

\begin{remark}
 Since there are exponentially many cycles in a graph in general, the \textsc{Cycle prevention condition~II} is hard to check. Especially in the case of non-linear consumption functions it can be already $\cal NP$-hard to check whether a specific cycle violates the inequality, since this also depends on the optimal choice of the $\alpha_{i}$.
\end{remark}

\section{Complexity Considerations}

For the case of constant consumption functions with $c_{b,e}\ge0$ for all edges, the problem under consideration transforms to the classical CSP, so the next theorem follows easily (see~\cite{Johnson_complexity}).

\begin{theorem} 
The problem $\bm{(P)}$ is $\mathcal{NP}$-complete.
\end{theorem}

Even when the \textsc{Cycle prevention condition~II} holds there are still exponential many different paths from $s$ to $t$ that have to be considered. Furthermore, even when considering only linear (or piecewise linear) consumption functions, problem $\bm{(P)}$ is already a mixed integer quadratically constrained quadratic program. For a given path (or cycle), it depends on the consumption functions whether one can find the optimal $\alpha_e$ easily. Nonlinear functions can make this task $\cal NP$-hard in the case of a nonconvex nonlinear program (see~\cite{Murty87}), whereas for linear consumption functions problem $\bm{(P)}$ transforms to a linear program of polynomial size.  

\begin{corollary}\label{coro:optalpha}
 For a fixed path $P$ and linear cost functions, the set of $a_i$ values for all $e_i\in P$ that minimize the fuel consumption on this path, can be determined in polynomial time.
\end{corollary}

This can, of course, be generalized to piecewise linear functions with a polynomial number of break points.
Furthermore, it is easy to check, whether the destination can be reached with pure electric drive, that is, for the case that $\sum_{e\in P} c_{f,e}(\alpha_e)=0 $. Since battery consumption is fixed ($\alpha=0$) in this case, the Bellman-Ford-algorithm can be use to compute such a path.

\begin{corollary}\label{coro:zeroconsumption}
 Deciding whether there exists a path with zero fuel consumption obeying the resource restrictions can be done in polynomial time.
\end{corollary}

\section{Constructing an FPTAS}\label{sec:approximation}

In this section we develop an FPTAS for our constrained shortest path problem $\bm{(P)}$ with subpath restrictions for battery consumption and control parameter $\alpha$. One main difficulty besides the choice of $\alpha_e$ is the feasibility of the path itself. Whereas it suffices to check feasibility only at the final node $t$ in the common CSP, we now have to check it at every intermediate point. Several algorithms for CSP, e.g., the approach of Ziegelmann~\cite{Ziegelmann_CSP}, make implicit use of this property by calculating a new weighted objective function containing also the resource bound. These ideas cannot be applied here. To achieve a {FPTAS} nevertheless, we apply the basic idea of Hassin's approach---rounding and scaling---in combination with the well known Bellman-Ford-algorithm applicable for networks with not necessary positive cost functions.

Let us fix source node $s$ and target node $t$. Furthermore, we are given an initial battery charge $B_0$ and upper bound $\overline{B}$, a non-negative cost function $c_{f,e} : [0,1] \to \R_{\geq 0}$, and a battery consumption function $c_{b,e} : [0,1] \to \R$ for each edge $e\in E$. Additionally, we require the \textsc{Cycle prevention conditions~I} and~\textsc{II} to be fulfilled. Thus, the optimal path has at most $|V|-1$ edges. Now, we want to find an $\varepsilon$-approximation of the shortest $s$-$t$-path with respect to $c_f$ obeying the battery constraints. That is, we want to find a feasible path that uses at most $(1+\varepsilon)$ times the costs of the optimal path. 

For each node, we introduce a set of labels, each label consisting of two values $(f,b)$ of fuel costs and remaining battery charge. These labels store the consumption values of the best subpaths found so far. To guarantee feasibility, the battery consumption has to be calculated exactly. Hence, in the approximation algorithm (following the Hassin approach) we are going to round the fuel values. Here, the main difficulty is to find the optimal precision for rounding. Choosing the precision too fine might result in a number of labels that cannot be bounded polynomially. Choosing it too coarse, we may not meet the approximation factor $\varepsilon$. 

Assume for the beginning that the minimum amount of fuel needed is already known and denote this value by \OPT. Now, we round up all occurring fuel values at the edges to integral multiples of $\frac{\varepsilon \OPT}{|V|-1}$. Thus, the error on each edge is at most $\frac{\varepsilon \OPT}{|V|-1}$. Since the optimal path  has at most $|V|-1$ edges, this limits the total error to at most ${\varepsilon \OPT}$. This yields at most $\frac{(|V|-1)(1+\varepsilon)}{\varepsilon}$ different fuel values that can occur at a node. Of course, for each node $v$ we only store the set $Q(v)$ of Pareto-optimal labels, i.e., the highest battery charge achieved so far for each of the possible fuel values. Initially, all label sets $Q(V)$ are empty, only node $s$ is labelled with $Q(s)=\{(0,B_0)\}$. 

In an update step, we propagate the whole label set of each node to all its neighbors. In detail, for every edge $e=(u,v)$ we compute all values of $\alpha_e$ such that $c_{f,e}(\alpha_e)$ yields the same discretization of possible fuel values. That is, we compute $\alpha_e^i$ with $c_{f,e}(\alpha_e^i)=i\frac{\varepsilon \OPT}{|V|-1}$ for all $i=0,\dots,\frac{(|V|-1)(1+\varepsilon)}{\varepsilon}$. Here, we assume that all operations concerning $c_f$ and $c_b$ can be computed in one time step. For each label $(f,b)\in Q(u)$, we calculate $(f+c_{f,e}(\alpha_e^i),b-c_{b,e}(\alpha_e^i))$ as possible new labels of $Q(v)$. The label set $Q(v)$ is updated accordingly if a new Pareto-optimal pair $(f,b)$ is identified. 

Since an optimal path can consist of at most $|V|-1$ edges, precisely  this number of propagation steps suffices to find an optimal $s$-$t$-path according to Bellman-Ford's algorithm. Thus, given any value $\OPT$, an $s$-$t$-path of length at most $(1+\varepsilon)\OPT$ can be found in time polynomial in $|V|$ and $\frac{1}{\varepsilon}$ if it exists.

Unfortunately, the \OPT\ value is unknown. Thus, we apply a binary search to find it. Obviously, $\LB=0$ is a lower bound on \OPT. Further, compute a shortest paths with respect to fuel consumption with $\alpha_e=\alpha_e^0$, i.e., the battery charge remains unchanged. Subsequently, compute optimal $\alpha_e$ for all edges of the obtained path with respect to the initial battery charge as described in Corollary~\ref{coro:optalpha}. The fuel consumption of this path is an upper bound $\UB$ on \OPT. Our initial guess on \OPT\ is simply $\overline{\OPT}=\frac{\LB+\UB}{2}$. 

If we find a feasible path with fuel consumption smaller than $\overline{\OPT}$, we can use this value as a new upper bound \UB. If we cannot find a path of length at most $(1+\varepsilon)\overline{\OPT}$, we can use $\overline{\OPT}$ as a new lower bound \LB. $\overline{\OPT}$ is updated accordingly and the binary search is stopped when $\frac{\UB-\LB}{\UB}$ is smaller than the required precision $\varepsilon$. If we also set pointers for each label to the preceding node where it originates from, we can reconstruct the path afterwards.

Note that this is just a brief and simplified sketch of the proposed algorithm. Several speed-ups are possible. If a lower bound $\LB>0$ is known, one may switch to a logarithmic scaling of the binary search, i.e., $\overline{\OPT}=\sqrt{\LB\UB}$. Further, one may already stop the binary search, if the gap is smaller than a certain predefined constant, say 2. Then one executes a final run with precision regarding $\LB$, but an enlarged label set up to $\UB$. Pointers for reconstructing the path have to be set only in this final run. Due to space limitations, we omit an in-depth runtime analysis here and refer the reader to the journal version of this paper.

\begin{theorem}
 For every $\varepsilon>0$, a path with fuel consumption at most $(1+\varepsilon)$ times the consumption of an optimal path can be found in time polynomial in the size of the input and $\frac{1}{\varepsilon}$.
\end{theorem}

If \textsc{Cycle prevention II} does not hold, the situation can change dramatically. A very short cycle (similar to the instance in Figure~\ref{fig:cycle2}) may occur very often in an optimal path. More precisely, the number of edges depends on the cost of a shortest cycle, i.e., the number of edges in an optimal path cannot be bounded polynomial in $|V|$. Even worse, if an edge is used multiple times in a path, rounding errors can become significantly larger. Thus, one may still apply the above algorithm to compute some path, but the approximation guarantee is lost.

\section{A more Practical Approach}

Although an FPTAS seems to be the best we can hope for regarding the complexity, the suggested approach bears some disadvantages. First of all, it is computationally expensive. With our application in mind, an efficient route should be computable by an on-board unit of the hybrid vehicle. Moreover, the required accuracy depends on the \OPT\ value itself, making pre-processing hardly usable.  Further, even pre-processing of the optimal control will only yield an approximation of the consumption functions. Thus, there is no need to aim at a higher accuracy in the routing than the optimal control can provide.

Given the graph $G=(V,E)$ together with its consumption functions, we construct a \emph{battery expanded network}\footnote{A similar idea is used for flows over time. Here, condensed time-expanded networks are use to approximate maximum flows (see~\cite{FleischerSkutella07}).}.
For each node $v\in V$, we add several copies $v_b$ each dedicated to a specific battery charge $b$. For example, we choose these battery values uniformly in the interval $[0,\bar B]$. Two nodes $u_{b_1}$ and $v_{b_2}$ are connected by an edge, if the original nodes were connected by an edge $e=(u,v)$. Now, we assign a constant fuel consumption to each edge $(u_{b_1},v_{b_2})$. This value matches the corresponding battery charges, i.e., we choose $\alpha$ such that $c_{b,e}(\alpha)=b_1-b_2$ and assign $c_{f,e}(\alpha)$ to the new edge. There may be no feasible choice of $\alpha$ for some edges, these edges are deleted. 

Now connect all copies $t_b$ to a new supertarget $t$ and add a new supersource $s$ connected to $s_b$ where $b$ corresponds to the largest value smaller than $B_0$. This yields a network with a positive cost function and without resource constraints. Any shortest path algorithm like Dijkstra or A* and several acceleration methods for these algorithms may be used to compute a shortest path from $s$ to $t$.

Assume \textsc{Cycle prevention condition~I} holds. For connecting two nodes of the same battery charge, we have to choose $\alpha_e=\alpha_e^0$. Thus, if there is a feasible path in the original network from $s$ to $t$, then there also exists a feasible path in the battery expanded network. However, it is more difficult to compute an approximation guarantee for the expanded network. Since the battery values are also rounded, the quality of the approximation depends on exchange ratios. If the error in battery charge can be expressed by means of the error in fuel, and this ratio is bounded, then also battery expanded networks can be used for constructing an FPTAS by appropriately choosing the battery charge levels.

This approach may also be used when \textsc{Cycle prevention condition~II} is not fulfilled. Now, an optimal path contains no cycles, but it may visit several copies of the same original node. This corresponds to cycles in the underlying unexpanded network, but now each cycle lifts the battery charge to a higher level. Consequently, the number of `cycles' in a shortest path depends on the granularity of the expansion.

\section{Conclusions and Future Work}

In this paper, we defined a model for routing hybrid vehicles, using two interconvertible resources in a constrained shortest path setting. We have shown that even with conservative cost functions an optimal path may contain cyles. We discussed assumptions on the cost functions that prevent such cycles and developed an FPTAS to find shortest paths. We also suggested a more practical approach based on a battery expanded network.

Our further research will focus on including travel times. For example, one may ask for a path, that is energy efficient but requires at most 10\% more travel time. Furthermore, we want to speed up the approximation. Here, one may think of an implicitly expanded network, that is, the algorithm works on the unexpanded network but necessary labels are handled like in the expanded network. Moreover, with a good heuristic one may not only propagate a single label but all labels of all copies of a node. Additionally, we want to improve the cycle prevention conditions. Here, it is very interesting whether one can find a condition where we can use some kind of cycle base to check it efficiently for all cycles.

This work was funded by the German Federal Ministry of Education and Research (BMBF), grant number 05M13ICA.

\appendix
\bibliography{arxiv_schwan_strehler_routing_of_hybrid_bibo}
\end{document}